\documentclass[letterpaper,twoside,11pt]{article}
\usepackage{amssymb}
\usepackage{amsmath}
\usepackage{latexsym}
\usepackage{verbatim}
\usepackage{graphicx}
\usepackage{epsfig}
\usepackage{pstricks}
\usepackage{rotating,graphicx}
\usepackage[english]{babel}
\usepackage{setspace}
\usepackage[english]{babel}
\usepackage[utf8]{inputenc}
\usepackage{natbib}
\usepackage[T1]{fontenc}
\usepackage{lmodern}
\usepackage{url}

\begin{document}


\begin{Large}
\begin{center}
\textbf{A Consciousness-Based Quantum Objective Collapse Model}\\[1cm]
\end{center}
\end{Large}

\begin{center}
\begin{large}
Elias Okon and Miguel Ángel Sebastián\footnote{This is a fully collaborative paper; authors appear in alphabetical order.} \\
\end{large}
\textit{Instituto de Investigaciones Filos\'oficas, Universidad Nacional Aut\'onoma de M\'exico, Mexico City, Mexico.} \\[.5cm]
\end{center}

\noindent 
Ever since the early days of quantum mechanics it has been suggested that consciousness could be linked to the collapse of the wave function. However, no detailed account of such an interplay is usually provided. In this paper we present an objective collapse model (a variation of the Continuous Spontaneous Location model) where the collapse operator depends on integrated information, which has been argued to measure consciousness. By doing so, we construct an empirically adequate scheme in which superpositions of conscious states are dynamically suppressed. Unlike other proposals in which ``consciousness causes the collapse of the wave function,''  our model is fully consistent with a materialistic view of the world and does not require the postulation of entities suspicious of laying outside of the quantum realm.

\onehalfspacing
\section{Introduction}
Quantum mechanics is often claimed to be the most successful physical theory ever constructed. It has an astonishing predictive power, continuously confirmed by empirical results. However, it remains controversial how the physical reality it describes is supposed to be portrayed. Particularly puzzling is the fact that quantum mechanics is committed to the \emph{superposition principle}, which holds that any two possible quantum states of a system can be added together to form another legitimate quantum state---a \emph{superposition} thereof. 

Although we can indeed observe the manifestation of such superpositions---as in the double-slit experiment---it seems there are at least some superpositions we fail to observe directly. The quantum formalism allows for the superposition of, say, a chair being in two different places, but we never come to observe such states. In standard presentations of the theory (e.g., \cite{dir,von}), this fact is accommodated by the collapse postulate, which states that, when we observe or measure a superposed state, like the one of the chair above, it collapses to one option or the other. The problem now turns into deciding when an \emph{observation} or \emph{measurement} happens, as this is not specified by the theory.\footnote{The standard theory not only does not specify when a measurement happens, it also does not prescribe what it is that is being measured (i.e., in which basis will the collapse occur).} 

It is natural to link the notion of observation with consciousness, an idea that has been floating in the air since the early days of quantum mechanics. \cite{von} argued that the mathematical formalism underlying quantum mechanics allows for the collapse of the wave function to be placed at any point in the causal chain between the measurement device and our ``subjective perception.'' Following this suggestion, \cite{LyB}  maintain that the collapse happens at the latter point in the causal chain, when consciousness takes place---an idea famously developed in \cite{Wigner}. Wigner holds that conscious states, unlike other states, do not admit superpositions. This in turn requires consciousness not to be a physical property or, at the very least, to be different from other physical properties in this respect. Other authors, such as \cite{CandM, Stapp:2005, Stapp:2007}, have more recently defended ideas along these lines.

Interpretations of quantum mechanics according to which ``consciousness causes the collapse of the wave function'' face several problems.\footnote{We use the term `cause' in a rather loose way just to honour the traditional motto of the idea that the collapse of the wave function depends somehow on consciousness. In our work there is no commitment whatsoever with the claim that there is such a thing as \emph{causation} at the fundamental level (see \cite{Price2007} for a discussion).} Many have found the commitment with mysterious non-physical entities unattractive. Moreover, if conscious states were indeed non-physical, then these theories would be committed to the existence of a mysterious interaction between consciousness and the physical world. This seems in conflict with certain basic principles of physics, such as the principle of energy conservation (cf. \cite{Averill:1981, Larmer:1986}), and would violate the common presupposition in scientific practices that the physical world is causally closed. On the other hand, if consciousness were something physical, it would remain mysterious how it would interact with the rest of the physical world and a description of the laws governing such an interaction would be forthcoming. The problem is that developing such dynamics is far from straightforward. 

For instance, it seems reasonable to assume (with Wigner) that consciousness does not admit superpositions. However, as has been recently pointed out in \cite{CandM}, this would be inconsistent with another reasonable assumption, namely, that collapses occur when systems get entangled with conscious beings (because if consciousness does not superpose, then it cannot get entangled). Moreover, if consciousness never superposes, and if the intervention of consciousness is what determines the precise moment at which the collapse of the wave function occurs, then one would run into trouble with the \emph{quantum Zeno effect}  (\cite{Dagasperis:1974}), which holds that frequent enough measurements effectively freeze the evolution of a system. Therefore, the effect would imply that it would not be possible for observed systems to evolve, and therefore, for our experience of the external world to change through time as it does.

In order to overcome some of these issues, still in the spirit of a theory in which consciousness is related to the collapse of the wave function, Chalmers and McQueen propose the introduction of what they call m-properties, whose superpositions are postulated either to be forbidden, or to be ``unstable'' or ``more likely to collapse.'' Moreover, they propose that some physical correlate of consciousness, such as integration of information (which Integrated Information Theory associates with consciousness, \cite{Tononi:2004}) could be an m-property. Note however that, in order for a theory that allows superpositions of consciousness to work, one would need to ensure for superpositions of different conscious states to quickly evolve into states of well-defined consciousness, in such a way that we would fail to notice these transitions in our experience. Moreover, it is key to recognize that, in order to have a satisfactory explanation of our observations, it is not sufficient to suppress superposition of conscious states, for this is still compatible with the existence of well-defined states of consciousness that correspond to the superposition of macroscopic properties---which we do not seem to experience. Therefore, a scheme along this lines would also have to somehow block these states.

In this paper we introduce a mathematically precise, consciousness-based collapse model that solves all the issues mentioned above, and which is compatible with the truth of materialism. In particular, we present an objective collapse scheme, a version of the Continuous Spontaneous Location (CSL) model, where the collapse operator is associated with a property that measures the level of consciousness. In particular, for illustrative purposes, we will employ integrated information---which, as we mentioned, has been argued to measure consciousness. By doing so, we arrive at a detailed proposal, in which superpositions of conscious states are dynamically suppressed in a way that is fully compatible with experience. As a result, we provide a way to fill a gap that consciousness-based interpretations have systematically left behind, greatly contributing to their unpopularity.

The rest of the paper is organized as follows. Section \ref{MP} introduces the measurement problem, briefly reviews the taxonomy of alternatives to address it and focuses on the objective collapse program. Then, with a consciousness-based model in mind, in section \ref{IIT} we explore the notion of integration of information, a concept that Integrated Information Theory associates with consciousness. We do not intend to offer a defense of such a theory, but we present it in some detail in order to motivate the relation between integration of information and consciousness. Finally, in section \ref{CSLC} we describe in detail our \emph{consciousness-based objective collapse model}, consider potential problems and explore ways to overcome them. In particular, we examine issues derived from the possibility of having well-defined states of consciousness that correspond to macroscopic superpositions and describe an \emph{evolutionary} solution to such a problem. To conclude, in section \ref{C} we briefly compare our model with standard collapse schemes and offer some closing remarks.

\section{The measurement problem}
\label{MP}
A common way to introduce the \emph{measurement problem} starts by calling attention to the fact that the standard formulation of quantum theory contains two very different evolution laws: the linear and deterministic Schrödinger equation and the non-linear and indeterministic collapse process, which interrupts the former during \emph{measurements}. The problem arises because, as John Stewart Bell convincingly argued in \cite{Bel:90}, higher-level notions, such as \emph{measurements}, should not appear as primitives in a theory that aims at being fundamental.

This and other complications with the collapse postulate have motivated attempts to do without it. However, such a move seems to lead to a related problem, often also referred to as the measurement problem. To see this, assume that \emph{everything}, \emph{always}  evolves according to the Schrödinger equation and consider a standard measurement scenario of, say, the spin along $z$ of a spin-1/2 particle. Let us look, then, at a measurement apparatus (which, as everything else, evolves according to the Schrödinger equation) that behaves as
\begin{equation}
|R\rangle_M |+\rangle_p  \xrightarrow{\text{Schrödinger}}  |+\rangle_M |+\rangle_p \quad \text{and} \quad
|R\rangle_M |-\rangle_p  \xrightarrow{\text{Schrödinger}}  |-\rangle_M |-\rangle_p ,
\end{equation}
where $|R\rangle_M$ is its ready state and with $|+\rangle_M$ and $|-\rangle_M$ the states where the apparatus displays spin-up and spin-down as the result of the experiment. In other words, the apparatus correctly tracks the $|+\rangle_p$ and $|-\rangle_p$ spin states of the particle and reacts accordingly.

Next, we ask what happens if the apparatus is fed with a particle in a \emph{superposition} of $|+\rangle_p$ and $|-\rangle_p$. Notice that the linearity of the Schrödinger equation implies
\begin{eqnarray}
\label{cat}
|R\rangle_M \left\lbrace \alpha |+\rangle_p + \beta |-\rangle_p \right\rbrace \xrightarrow{\text{Schrödinger}} \alpha |+\rangle_M|+\rangle_p + \beta |-\rangle_M|-\rangle_p ,
\end{eqnarray}
which means that the apparatus itself will end up in a superposition of displaying spin-up and spin-down. Such a state seems at odds with experience. To drive the point home, we can introduce an observer (which is also assumed to evolve according to the Schrödinger equation) that looks at the apparatus during the measurement. The linearity of the Schrödinger equation, once more, leads to a superposition, this time of the observer \emph{perceiving} spin-up and spin-down as the result of the experiment. That is
\begin{equation}
\label{SP}
|R\rangle_O |R\rangle_M \left\lbrace \alpha |+\rangle_p + \beta |-\rangle_p \right\rbrace \xrightarrow{\text{Schrödinger}} \alpha |+\rangle_O |+\rangle_M|+\rangle_p + \beta |-\rangle_O |-\rangle_M|-\rangle_p ,
\end{equation}
where $|R\rangle_O$ is the state of the observer in which she is ready to look at the result displayed by the apparatus and $|+\rangle_O$ and $|-\rangle_O$ are the states in which she perceives that the results where spin-up and spin-down, respectively (we are making the very reasonable assumption that the state of the observer correctly tracks the state of the macroscopic system). The problem with all this, of course, is that the final state in equation (\ref{SP}), which describes a \emph{superposition of perceptions}, does not seem to correspond with what we experience when we perform such an experiment; or, as David Albert puts it in \cite[p. 78-79]{Alb:92}, such a state ``is at odds with what we know of ourselves by \emph{direct introspection}.'' 

Note however that this strange discrepancy between the predictions of a theory with pure Schrödinger evolution on the one hand, and our experience on the other, crucially depends on \emph{interpreting} the quantum state along the lines of the so-called Eigenvector-Eigenvalue (EE) link. Such a rule states that a physical system possesses the value $\alpha$ for a property represented by the operator $O$ if and only if the quantum state assigned to the system is an eigenstate of $O$ with eigenvalue $\alpha$. In other words, if a state is represented by a given vector and such a vector is an eigenstate of the operator that represents a property, then we can claim that the system possesses a defined value for such a property and that its value is the corresponding eigenvector. Likewise, if a system possesses a defined value for some property, then, the vector which represents its state must be an eigenstate of the operator which represents such a property. The point is that only if one interprets quantum states along these lines, one has to read the final state in equation (\ref{SP}) as one in which the observer does not have a well-defined perception. It follows that if one interprets the quantum state differently, one could escape the measurement problem. 

One such an alternative interpretation, first introduced by Hugh Everett III in \cite{Eve:57}, proposes to read the final state in equation (\ref{SP}) not as a state in which the observer does not have a well-defined perception, but one in which the observer simultaneously, but \emph{independently}, has both perceptions. That is, one in which the observer, as Michael Lockwood puts it in \cite[p. 166]{Loc:96} ``is \emph{literally} in two minds''. It is very important to notice, though, that an interpretation along the lines of the Eigenvector-Eigenvalue link is crucial in the process of extracting concrete predictions from the theory (because it is such a tool that allows to connect statements regarding what the quantum state is with statements regarding what obtains in the world). Therefore, if one changes the interpretation, one has to make sure that the alternative one also leads to empirically successful predictions---something not at all clear under the interpretation suggested by Everett (see, e.g., \cite[sec. 4]{MW}).

Alternatively, one could escape interpreting the final state in equation (\ref{SP}) as one in which the observer does not have a well-defined perception by adding extra elements to the picture---such as Bohmian particles---that determine which of the two terms of the superposition actually obtains in the world (see \cite{Bohm}). This is certainly a promising path, but in this work we will focus on alternatives that assume that the quantum state is complete.

Going back to the measurement problem, it is one thing to try to pinpoint exactly what it consists of and another to be clear about what a satisfactory solution to the problem looks like. Regarding the latter, probably a far more useful task than the former, one could say that a satisfactory solution to the measurement problem consists of a formalism which:
\begin{enumerate}
\item Is fully formulated in precise, mathematical terms (with notions such as \emph{measurement}, \emph{observer} or \emph{macroscopic} not being part of the fundamental language of the theory).
\item Reproduces the empirical success of standard quantum mechanics at the microscopic level.
\item Explains why certain macroscopic superpositions allowed by the theory never seem to occur.
\end{enumerate}

The last point needs a bit of unpacking. First of all, it is very important to highlight an often overlooked distinction between a \emph{superposition of perceptions} and a \emph{perception of a superposition}. That is, between:
\begin{itemize}
\item A superposition of incompatible perceptions, such as the final state in equation (\ref{SP}): $\alpha |+\rangle_O |+\rangle_M|+\rangle_p + \beta |-\rangle_O |-\rangle_M|-\rangle_p$.
\item A well-defined perception of a macroscopic system in a superposition of different positions: $|S\rangle_O \left\lbrace \alpha |+\rangle_M|+\rangle_p + \beta |-\rangle_M|-\rangle_p \right\rbrace = |S\rangle_O |S\rangle_{M+p}$ (where $|S\rangle_{M+p} \equiv \alpha |+\rangle_M|+\rangle_p + \beta |-\rangle_M|-\rangle_p $ and $|S\rangle_O$ corresponds to a state in which the observer experiences a well-defined perception of the measurement apparatus displaying a superposition of spin-up and spin-down).
\end{itemize}
Regarding the first scenario, we have already explained how it may arise and why it represents a problem. Regarding the second, the point is that, from the \emph{quantum} point of view, both $|S\rangle_{M+p}$ and $|S\rangle_O$ are states which are on a par with states such as $|+\rangle_O$ or $|+\rangle_M|+\rangle_p$, and the only reason they may seems strange, different or unacceptable is because we in fact do not experience them. The key question, of course, is why? One could  question whether $|S\rangle_O$ is in fact a state of well-defined perception. Well, maybe it is not, and we will have more to say about it later, but what is clear is that $|S\rangle_O$ is not a superposition of other states of well-defined perception, such as $\alpha |+\rangle_O  + \beta |-\rangle_O $, and that, if it turns out not to be a state of well-defined perception, an explanation of such a fact would be required.

To sum up, states such as $|S\rangle_O$ or $|S\rangle_{M+p}$ are perfectly valid from the fundamental, quantum point of view and a satisfactory quantum formalism is required to explain, in precise mathematical terms, why no one has ever reported experiencing states of that kind. That is, we never seem to experience well-defined conscious states that correspond to a superposition of external macroscopic objects, such as chairs, being in two positions. It is in fact common to find two different explanations for this, both having to do with decoherence and both problematic.\footnote{Decoherence studies the consequences of the inevitable interaction between a quantum system and its macroscopic environment.}  The first one asserts that we cannot be sure that we never experience states such as $|S\rangle_O$ because, for that, we would need to perform an interference experiment with a macroscopic object and decoherence gets in the way. The problem with such an answer is that it equivocates between a superposition of perceptions and a perception of a superposition (see bullets above). The point is that some type of an Everettian move plus decoherence could probably explain why a state such as the final one on equation (\ref{SP}) is perceived as both terms of the superposition independently, but it has absolutely nothing to say about why would something similar happen with $|S\rangle_O$. The second explanation often offered for why no one has ever reported experiencing states such as $|S\rangle_{M+p}$ is that, because of decoherence, they are somehow dynamically suppressed. This is something that is nowadays extremely common to read or hear, the problem is that it never comes with adequate backing (see \cite[sec. 2]{Oko.Sud:16}). At any rate, we will have more to say about this below.

Returning to the issue of finding a satisfactory quantum theory, it is clear that the standard interpretation of quantum mechanics does not qualify as one because it does not meet point 1 of the list: collapses occur upon \emph{measurements}. However, as we will see in the next subsection, there are more acceptable ways to introduce collapses into the picture. 

\subsection{Objective collapse models}
With the measurement problem in mind, \emph{objective collapse} (or \emph{dynamical reduction}) models aim at constructing a single dynamical equation that adequately encompasses both the standard unitary evolution and the collapse mechanism. The idea is to add non-linear, stochastic terms to the Schrödinger equation in such a way that the behavior at the microscopic level is not significantly altered (with respect to the standard framework), but where embarrassing macroscopic superpositions are effectively suppressed.

In the simplest collapse model, know as GRW (\cite{GRW}), all elementary particles are postulated to suffer, with mean frequency $\lambda_{GRW}$, spontaneous localization events around randomly chosen positions. The localizations are implemented by a multiplication of the wave function by narrow Gaussians, with centers selected according to a probability distribution that mimics the Born rule. Given that the collapse frequency of an object is proportional to the number of particles it contains, a macroscopic object will be highly likely to suffer collapses even if $\lambda_{GRW}$ is extremely small (which is needed for the microscopic behavior not to be affected significantly). Moreover, given that it is enough for one of the particles of a macroscopic superposition (such as that in the final state in equation (\ref{SP})) to get localized, for the whole state to collapse, the GRW model ensures a quick elimination of superpositions of well-localized macroscopic states, with statistics in accordance with the standard theory.

The Continuous Spontaneous Localization model, or CSL (\cite{CSL}), replaces the discontinuous GRW jumps with a continuous, stochastic evolution equation. In more detail, it adds specific non-linear, stochastic terms to the Schrödinger equation designed to drive any initial wave function into one of the eigenstate of a, so-called, collapse operator. In the simplest case, the solutions to the CSL equation are given by 
\begin{equation}
\label{CSL}
|\psi(t)\rangle_B = e^{- \left\lbrace iH+\frac{1}{4\lambda t} \left[ B (t) - 2\lambda \hat{A} \right]^2 \right\rbrace } |\psi(0)\rangle
\end{equation}
with $\lambda$ a free parameter that controls the strength of the stochastic terms, $\hat{A}$ the collapse operator and $B(t)$ a classical Brownian motion function selected randomly with probability density
\begin{equation}
\mathcal{P}_t \{B\}  = {}_B\langle\psi(t)|\psi(t)\rangle_B .
\end{equation} 
Note that the first term in the exponent corresponds to the standard Schrödinger evolution and the rest are the additional stochastic terms that implement the collapse. 

To see how the model works, we expand the initial state at the right-hand-side of equation (\ref{CSL}) in terms of a superposition of eigenstates of $\hat{A}$
\begin{equation}
|\psi(0)\rangle = \sum_i c_i |a_i\rangle
\end{equation}
and, taking for simplicity $H=0$, we arrive at 
\begin{equation}
\label{E1}
|\psi(t)\rangle_B = \sum_i c_i e^{- \frac{1}{4 \lambda t} \left[ B (t) - 2\lambda t a_i \right]^2} |a_i\rangle
\end{equation}
and
\begin{equation}
\label{E2}
\mathcal{P}_t \{B\}  = \sum_i e^{- \frac{1}{2 \lambda} \left[ B (t) - 2\lambda t a_i \right]^2} |c_i|^2.
\end{equation} 
From the last equation we note that the most probable $B(t)$'s to occur are $B(t) \approx  2 \lambda t a_j$ with probabilities $|c_i|^2$, in which case
\begin{equation}
\label{E3}
|\psi(t)\rangle_B \approx  c_j |a_j\rangle + \sum_{i \neq j} e^{- 2 \lambda t \left[ a_i - a_j \right]^2} |a_i\rangle \xrightarrow{t \rightarrow \infty}  c_j |a_j\rangle.
\end{equation}
Therefore, as $t \to \infty$, the CSL dynamics drives the state of the system into the $j$-th eigenstate of the operator $\hat{A}$, with probability $|c_j|^2$; that is, it unifies the standard unitary evolution with a ``measurement'' of such an observable.

Given the nature of these models, one could worry for them to suffer a problem similar to the Zeno effect, in which systems get forever frozen on eigenstates of the collapse operator. Note however that the \emph{full} time-evolution under these models contains both the standard unitary component (first term in the exponent of equation (\ref{CSL})) and the collapse mechanism (second term of the exponent of equation (\ref{CSL})), so the actual evolution of a system involves a competition between the two. Of course, the result of this struggle is to be decided by the strength of the collapse terms, which is determined by the parameter $\lambda$, so the key question is if there is a possible value for $\lambda$ that avoids these problems and yields empirically successful predictions.

More generally, the value of the collapse parameter has to satisfy several constraints. On the one hand, $\lambda$ cannot be too large; otherwise, microscopic phenomena, which we know are well-described by a purely unitary evolution, would get disturbed. Moreover, a large $\lambda$ would lead to the Zeno-type problem in which the collapse terms would dominate and eigenstates of the collapse operator would freeze. On the other hand, if $\lambda$ is too small, these models would not achieve their purpose of suppressing undesirable macroscopic superpositions. Of course, one can allow for these macroscopic superpositions to persist for some time, but one need to make sure for them to quickly die-out before we are able to notice them. Well, the beauty of these collapse models is that there exists a possible range of values for $\lambda$ that yields the required equilibrium, i.e., that provides us with fully empirically successful models of the world around us (see \cite{Adl,FandT,Bas:13}.\footnote{A more serious complication regarding collapse models arises from the fact that they lead systems to states which are very close to eigenstates of the collapse operator, but not exactly to such eigenstates. Therefore, if one subscribes to the EE link, systems under collapse dynamics never actually possess well-defined values for properties associated with the collapse operator (nor for most other properties). The solution, then, is to substitute the EE link by something else. One alternative is the, so-called, \emph{fuzzy link} interpretation introduced in \cite{AandL}, in which one allows for some tolerance away from an eigenstate while ascribing the possession of well-defined properties. Another alternative is to construct out of the wave function a, so-called, primitive ontology, such as mass density or flashes, and to interpret such an entity as the three-dimensional stuff that populates the world (see \cite{All}). It is fair to say, though, that these approaches, while promising, still have some open issues to address (see, e.g., \cite{Mc}).}

Going back to CSL, note than, unlike GRW, which is by construction associated with the position basis, it allows the freedom to select different collapse operators. However, $\hat{A}$ is usually chosen to be associated with the position operator. That is because, as in GRW, such an option leads to the suppression of superpositions of macroscopic objects at different locations, and thus to a solution to the measurement problem. In fact, it has even been argued that this is the only option available (see \cite{Bas:03}). However, we will show that a very different choice for a collapse operator can also lead to a solution of the measurement problem. 

The point is that, in order to explain why we never perceive superposed macroscopic objects, at least two options are available: one can construct models in which such macroscopic superpositions never occur (as in standard collapse theories) or one can maintain that, although such superpositions do occur, we never encounter them because they collapse as soon as we observe them. Below we will explore this second group of alternatives and present a version of CSL in which the collapse operator relates to consciousness. This of course requires the construction of an operator that measures consciousness and, for this purpose, we will follow Integrated Information Theory and employ a measure of integration of information. In order to motivate the relation between integration of information and consciousness, in the next section we present such a theory in some detail.

\section{Integration of Information and Consciousness}
\label{IIT}
Integrated Information Theory (henceforth IIT) is a novel theory of consciousness proposed in 2004 by Giulio Tononi (\cite{Tononi:2004,Tononi:2008,Tononi:IIT3:2014}). The theory has gained popularity in recent years, especially in neuroscience, where it has even attracted the attention of some renowned neuroscientists in the field of consciousness studies, such as Christopher Koch (see, e.g.,  \cite{Tononi:Koch:2015}).

The core idea behind IIT is that consciousness, at the fundamental level, is \emph{integrated information}, which is described as ``the amount of information generated by a complex of elements, above and beyond the information generated by its parts'' \cite[p. 216]{Tononi:2008} or ``information specified by a whole that cannot be reduced to that specified by its parts''\cite[p.1]{Tononi:IIT3:2014}.\footnote{In its latest formulation, what has been called IIT 3.0, the proponents of IIT depart slightly from Shannon's notion of information, which they call `extrinsic information,' and focus in what they call `intrinsic information.'} In order to quantify integrated information, IIT defines the property $\Phi$, which is taken to measure  the quantity of consciousness generated by a complex of elements. The quality of experience (the particular experience we undergo) is then associated with the set of informational relationships generated within that complex.

Tononi and colleagues derive these ideas from a set of axioms they take to be self-evident. From those axioms they deduce  what they call postulates, which are supposed to specify conditions a system must satisfy in order to generate consciousness. For current purposes, we can focus on three axioms that primarily motivate $\Phi$ as a measure of consciousness, Information, Integration and Exclusion \cite[pp.2-3]{Tononi:IIT3:2014}:
\begin{description}
\item[INFORMATION:] Consciousness is informative: each experience differs in its particular way from other possible experiences. Thus, an experience of pure darkness is what it is by differing, in its particular way, from an immense number of other possible experiences. A small subset of these possible experiences includes, for example, all the frames of all possible movies. 
\item[INTEGRATION:] Consciousness is integrated: each experience is (strongly) irreducible to non-interdependent components. Thus, experiencing the word ``SONO'' written in the middle of a blank page is irreducible to an experience of the word `` SO'' at the right border of a half-page, plus an experience of the word ``NO'' on the left border of another half page the experience is whole. Similarly, seeing a red triangle is irreducible to seeing a triangle but no red color, plus a red patch but no triangle.
\item[EXCLUSION:] Consciousness is exclusive: each experience excludes all others – at any given time there is only one experience having its full content, rather than a superposition of multiple partial experiences; each experience has definite borders – certain things can be experienced and others cannot; each experience has a particular spatial and temporal grain.
\end{description}

Some readers might not find these axioms as self-evident as Tononi and colleagues pretend. More usefully, \cite{Tononi:2008} makes use of thought experiments to make the relation between $\Phi$ and consciousness more compelling. Imagine you are participating in an experiment where you are facing a screen with a light that can be either \emph{on} or \emph{off}. Your task is to say ``on'' if the light is \emph{on} and ``off'' otherwise. The same task can be performed by a photodiode, which can discriminate between the \emph{on} and \emph{off} states by changing the output current. The photodiode can then be connected to a device that says ``on'' if the current is above certain threshold and ``off'' otherwise. Although the photodiode can make the same discrimination than you, we would not think it has any subjective experience of the process. Tononi asks what the difference is, and replies that it lies on how much information is generated when the distinction is made. Classically, information is defined as a reduction of uncertainty and measured by means of Shannon's entropy, which is roughly given by the logarithm of the number of alternatives. Therefore, according to Tononi, the main difference between the photodiode and us lies on the restriction in possibilities that our states make in comparison with those of the photodiode. By being in an ``on'' state, the ``off'' state of the photodiode is ruled out. In contrast, our experience also rules out the light being located in a different location, being of a different color, etc. The  key point, according to Tononi, is to realize how ``the many discriminations we can do, and the photodiode cannot, affect the meaning of the discrimination at hand, the one between light and dark'' \cite[p. 218]{Tononi:2008}.

The second thought experiment has to do with integration. In this case we are asked to consider a digital camera, which contains, say, a million photodiodes like the one in the previous example. Therefore, the camera can distinguish $2^{1000000}$ different possibilities and, clearly, we could increase the number of sensors in such a way that the camera discriminates as many alternatives as we visually do. Yet, few would be willing to claim that the camera is conscious. What is the difference in this case? Tononi proposes that it is the way in which we do the discrimination. We, contrary to the camera, do it as an integrated system: ``one that cannot be broken down into independent components each with its own separate repertoire'' \cite[p. 219]{Tononi:2008}. Whereas the information of the camera can be perfectly analyzed in terms of the information of the sensors, we cannot do this in the case of our experience, which is always integrated (i.e., from the point of view of the experience, say, of a red triangle, it does not make sense to take apart the experience of red from the experience of a triangle).

As we said above, in order to quantify the amount of integrated information in a system, IIT introduces the property $\Phi$. Although the precise mathematical calculation for $\Phi$ has been modified over the years, the core idea, quantifying information in a system, above and beyond that of its parts, remains the same. It is important to mention that, within the theory, the partition of a system into separate components or mechanisms depends on the selected level of abstraction and spatio-temporal scale. Moreover, IIT does not specify the spatio-temporal scale at which systems have to be considered. Thereofre,  $\Phi$ has to be calculated for all possible partitions of the system, at every spatio-temporal grain, and the level that locally maximizes $\Phi$, attaining value $\Phi^{Max}$, is the one responsible for consciousness in the system (e.g., \cite{Hoel:Tononi:2013} show that $\Phi$ can peak at a macroscopic, rather than microscopic, spatio-temporal scale).\footnote{Calculating $\Phi^{Max}$ might very well be beyond our cognitive capacities. However, for current purposes, the only thing required is that, for every state of a system, there is a well-defined value of $\Phi^{Max}$, independently of our capacity to come to know it.} 

Tononi and colleagues have focused on a simplified calculus, in which they model a system as a logic gate network (see \cite{Tononi:2008} for a detailed explanation of the calculus within such a model). The idea is that logic gates offer a plausible approximation, if the neural level is the adequate one. It is important to note that if we want to fix the behavioral disposition of an organism, and if computationalism is correct, then it suffices to fix the computations as described at the computational level, independently of the way in which such a computation is realized. That is, two systems might differ with respect to the mechanisms involved at a certain level or spatio-temporal scale (e.g., at the fundamental level), without differing in its mechanisms at another level (e.g., the computational one).

Summing up, the property $\Phi^{Max}$ is supposed to quantify the amount of consciousness in a system. In the next section, we explore the quantity $\Phi^{Max}$ at the quantum level---which we assume to be the fundamental one---and we use it to construct a consciousness-based version of CSL. Our model is intended to illustrate, in a conceptually and mathematically sound manner, how consciousness could play a role in the collapse of the wave function.\footnote{The model that we present is consistent with different metaphysical views regarding the  details of the relation between maximally integrated information and consciousness. For example, one might endorse something along the lines of IIT and think, as a referee has suggested to us, that maximal integrated information is ``the mark'' or correlate of consciousness, whereas we still need some underlying (proto-) phenomenal property at the fundamental level that accounts for the Hard Problem \citep{chalmers:CM:1996}. Whether panpsychism is in a better position to account for (a version of the) Hard Problem is an open question---see \cite{ChalmersCombination}, \cite{Goff2009}, \cite{Sebastian_panp_2015}. For discussion of the compatibility between panpsychism and IIT see \cite{Morch2018}. 

Alternatively, one could claim that states with maximally integrated information are a posteriori identified with phenomenally conscious states; i.e. acknowledge the lack of a priori explanation of consciousness and hold that materialism can be true a posteriori (for discussion see \cite{Chalmers:two-dimensional:2009}). 

Although some proponents of IIT are happy to accept some panpsychist consequences, it is unclear that they do so for reasons related to the Hard Problem. In fact, Tononi and colleagues stress that they take a completely different route with regard to the hard problem, which they take their proposal to address: 
``IIT addresses the hard problem in a new way. It does not start from the brain and ask how it could give rise to experience; instead, it starts from the essential phenomenal properties of experience, or axioms, and infers postulates about the characteristics that are required of its physical substrate.''\citep{Tononi2016}

For the purpose of the paper we remain neutral with regard to the relation between IIT and the Hard Problem, as well as between the relation between the Hard Problem and materialism.} 

\section{A consciousness-based CSL model}
\label{CSLC}
In this section we lay out the details of our proposal. In a few words, it consists of a CSL model with $\Phi^{Max}$ as the collapse operator. By doing so, we arrive at a model in which, as has been suggested throughout the years, consciousness plays a role in the collapse of the wave function. The advantage of our proposal, of course, is that we incorporate consciousness into quantum theory in a perfectly well-defined way, both mathematically and conceptually.

Before presenting our proposal it is important to mention that, in \cite{Kre.Ran:15}, a CSL model in which $\Phi^{Max}$ plays a role is also constructed. However, in such a work $\Phi^{Max}$ is associated with the rate of collapse and not with the collapse operator (which is, as in standard CSL, taken to be related to the position basis). That is, in such a model, as in standard CSL, the collapse basis is  associated with the position operator, but the rate of collapse $\lambda$ is taken to be a (monotonically increasing) function of $\Phi^{Max}$. The idea is that the presence of consciousness (via a large value of $\Phi^{Max}$) amplifies the chance of a collapse. There is, however, a defeating problem for current purposes with such a proposal, for it assumes that the value of $\Phi^{Max}$ is always well-defined. Otherwise, the rate of collapse would not be  well-defined at all times. The problem, of course, is that in the quantum context under consideration, $\Phi^{Max}$, as any other property, cannot always posses a well-defined value. In reply, one could  assume that $\Phi^{Max}$ behaves differently, but this undermined the objective of constructing a consciousness-based quantum theory without the introduction of extraneous entities.

The first issue to discuss is the construction of a quantum version of $\Phi^{Max}$. To begin with, we remember that, according to quantum theory, to every property of a system corresponds a Hermitian operator. Given that, as we explained above, $\Phi^{Max}$ is a well-defined property of any system, then there must be a corresponding operator $\hat{\Phi}^{Max}$ that represents such a property. The idea, then, is that only states with a well-defined value of $\hat{\Phi}^{Max}$ can be conscious. That is, only eigenstates of $\hat{\Phi}^{Max}$ correspond to conscious states. Next comes the question of how to define $\hat{\Phi}^{Max}$. Given that it is supposed to measure how much information a system contains, \emph{above and beyond that of its parts}, and given that \emph{quantum entanglement} is precisely related to such an issue, it is natural to define $\hat{\Phi}^{Max}$ in terms of some measure of entanglement. The standard measure of entanglement for pure states is the entanglement entropy, but such a measure is no longer useful for mixed states. For the latter there are a number of options, such as entanglement cost, distillable entanglement, entanglement of formation, relative entropy, squashed entanglement or logarithmic negativity, but none of them is standard (see \cite{entang} for a review). In \cite{Kre.Ran:15}, for instance, $\hat{\Phi}^{Max}$ is defined in terms of \emph{relative entropy}.

Finally, there is the issue of, in order to calculate $\Phi^{Max}$ for a given system, having to calculate $\Phi$ for all of its parts, for all of its partitions and  all possible levels of description.\footnote{Here we are assuming a physicalist position according to which all possible levels of description of a system, e.g., chemical, biological, economical, etc., are, at the end of the day, already present (albeit in an extremely complicated way) in the fundamental, quantum mechanical description.} Therefore, given a quantum system, in order to calculate $\hat{\Phi}^{Max}$ it is necessary to consider all its subsystems, all possible partitions of each of those subsystem and also different descriptions of the system at all possible coarse-grainings. Needless to say, doing all this for a realistic system is beyond present technical abilities. What is important for us, though, is that the operator $\hat{\Phi}^{Max}$ is guaranteed to exist.

Putting all together, for a given initial state $|\psi(0)\rangle$, the CSL model we propose has as solutions
\begin{equation}
\label{CSLPhi}
|\psi(t)\rangle_B = e^{- \left\lbrace iH+\frac{1}{4\lambda t} \left[ B (t) - 2\lambda \hat{\Phi}^{Max} \right]^2 \right\rbrace } |\psi(0)\rangle
\end{equation}
 with $B(t)$ a classical Brownian motion function selected randomly with probability density
\begin{equation}
\mathcal{P}_t \{B\}  = {}_B\langle\psi(t)|\psi(t)\rangle_B .
\end{equation} 
Therefore, our model is such that it drives any initial state of a system into an eigenstate of the $\hat{\Phi}^{Max}$ operator. If well-defined values of $\hat{\Phi}^{Max}$ are indeed related to conscious states, then, in the same way that standard CSL quickly destroys  Schrödinger cat states, the above dynamics quickly kills superpositions of incompatible conscious states, leading to states of well-defined consciousness.

There is a potentially serious objection to the claim that our model effectively suppresses superpositions of conscious states, particularly in measurement scenarios.\footnote{We thank an anonymous referee for rising this interesting objection.} On the one hand, CSL models do not collapse superpositions of eigenstates that possess the same eigenvalue.\footnote{This can be seen by noting the role eigenvalues play in equations (\ref{E1}), (\ref{E2}) and (\ref{E3}).} On the other hand, it might seem that the eigenvalues of $\hat{\Phi}^{Max}$, corresponding to two different outcomes of a typical measurement, will be equal. That is because the integrated information for those two states seems to be the same (e.g. both outcomes might be registered in different regions of the brain, but the registering structures will be the same). If these eigenvalues are indeed equal, then our model, after all, will not suppresses this type of superpositions of conscious states. 

However, on second thought, it is extremely unlikely for two eigenvalues of $\hat{\Phi}^{Max}$, corresponding to two different outcomes of a measurement, to be the equal. This is because, as we learn from decoherence, any quantum system inevitably interacts with its environment, and by doing so, it gets entangled with it in such a way that the states of the environment corresponding to different outcomes of a measurement are almost orthogonal---and hence very different among themselves. Given that the conscious state of the observer is influenced not only by the result of the measurement, but also by its environment, its states corresponding to different outcomes will not only differ on the value of the result, but also on the myriad of changes to the environment brought about by the fact that a particular value was obtained. As a result, in general, the values of integrated information corresponding to different results of the experiment will be different, leading in our model to an effective suppression of superpositions of conscious states. One could argue that responding to this objection by appealing to an interaction with the environment would imply that consciousness is left out of the picture, but this is not the case. Of course, the state of the environment is relevant, but only insofar as its effect on the observer is. That is, within our model, it is only because conscious systems react differently to different environments that superpositions of conscious states collapse. 

Looking back at the list of necessary components of a satisfactory solution to the measurement problem presented  in section \ref{MP}, we see that, so far, our model seems promising. To begin with, it is fully formulated in precise, mathematical terms and, as long as the CSL parameter is small enough, it reduces to standard quantum mechanics at the microscopic level. Regarding an explanation of why we never seem to encounter certain macroscopic superpositions allowed by the standard theory, above we made an important distinction between two different scenarios to explain away: i) superpositions of incompatible perceptions and ii) well-defined perceptions of a macroscopic system in a superposition of different positions. It is clear that our model takes care of the first complication by not letting those states last for long. As with standard collapse schemes, one could worry that our model could lead either to a Zeno-type problem, in which conscious states freeze, or to superpositions of conscious states that we could actually experience. However, the same value for the collapse parameter $\lambda$ that allows for the construction of empirically successful standard collapse models, would also do the trick here.\footnote{A related worry is that it would be hard for consciousness to evolve in the early universe because the collapse mechanics would freeze it at an eigenstate with $\Phi^{Max}=0$. However, the situation, again, is completely analogous with standard, position-based CSL, which of course can be applied to the early universe without implying that the universe would freeze at some eigenstate of position and nothing would ever move (see \cite{P7,P9,P14} for successful applications of standard CSL to the early universe).}

One might find puzzling the idea of there being superpositions of conscious states that we fail to notice. However, it is easy to make sense of the distinction between the experience we have and what we notice, by means of the conceptual distinction between consciousness and cognitive access---or between phenomenal consciousness and access consciousness \citep{Block2002}. The point is that the things we come to ``notice’’ depend upon the availability of information for thought. Although we often say, in ordinary language, that we are conscious of such and such, just in case we ``notice’’ such and such, the term `consciousness’ in the discussion refers exclusively to our subjective experience. It is then an open empirical question whether the mechanisms that underlie our subjective experience depend upon those that make the information available for thought. And there is strong empirical evidence---endorsed by the proponents of IIT \citep{Tononi:Koch:2015}---suggesting a response in the negative \citep{Block:2011,Block:2014,Sebastian:synthese2014}. In this case, just as there are changes in the external world we miss because they are too quick for our perceptual system to register them,\footnote{What ``too quick'' means in this case depends on the luminance \citep{Bloch, Scharnowski:2007}.} if the time during which our conscious states enter into a superposition is short enough---something that depends, as we have seen, upon the value of the collapse parameter $\lambda$---we would miss that as they are too quick for the corresponding access-process to register them: for us to notice them.\footnote{One may also worry about the fact that our model does not really lead systems to eigenstates of $\hat{\Phi}^{Max}$, but only to states which are very close to those eigenstates. As with standard collapse models, if one strictly follows the EE link one gets into trouble because you would have to conclude that our model leads to a scenario in which conscious states never actually occur. The solution, again, as with standard collapse models, is to deviate from the EE link and introduce some type of fuzzy link that ascribes consciousness to states which are close enough to $\hat{\Phi}^{Max}$ eigenstates. How to define such a ``close enough'' is still an open question.}

Above we mentioned that there are two different scenarios involving macroscopic superpositions that we need to explain away and we explained how our model takes care of the first one, namely, superpositions of different perceptions. As we will see below, dealing with the second one, i.e., well-defined perceptions of macroscopic superpositions, is more complicated. The problem is that states such as $ |S\rangle_O |S\rangle_{M+p}$ are \emph{not} suppressed by the model. What we need, then, is a way of restricting the number of accessible states. Below we will propose an \emph{evolutionary} explanation for this restriction.

\subsection{Well-defined perceptions of a macroscopic superposition}
Suppose an observer measures the spin of a particle. In order for our model to correctly predict the fact that she will end up either observing spin-up or spin-down, one has to further assume that the states $|+\rangle_O$ and $|-\rangle_O$ are eigenstates of $\hat{\Phi}^{Max}$. The problem is that it could very well be that this is not the case and that the eigenstates of $\hat{\Phi}^{Max}$ are superpositions of such states. That is, there is nothing in the theory to select such a basis over others. It is an empirical fact, however, that when we perform this type of experiments we end up with perceptions corresponding to $|+\rangle_O$ or $|-\rangle_O$ so it seems to be the architecture of our brain that selects such a basis. The point however is that once we assume that $|+\rangle_O$ and $|-\rangle_O$ are eigenstates of $\hat{\Phi}^{Max}$, it follows that superpositions thereof are not (assuming of course no degeneracy). As we explained above, though, this does not mean that states such as $|S\rangle_O$ are superpositions of states such as $|+\rangle_O$ and $|-\rangle_O$. As we said, $|S\rangle_O$ is, by assumption, an eigenstate of $\hat{\Phi}^{Max}$ corresponding to a well-defined perception of a macroscopic superposition and nothing we have said so far explains why we do not seem to experience such states.

As we mentioned in section \ref{MP}, a common explanation for the fact that we never seem to find ourselves in states such as $|S\rangle_O$ is that they are dynamically suppressed.  However, the fact that, from the fundamental point of view, $|S\rangle_O$ and, e.g., $|+\rangle_O$, are on an equal footing, makes it hard to see where the equivalence between them could break. Of course, as is well-known, decoherence is supposed to come to the rescue. The idea is that a preferred basis is selected by the fact that all states inevitably interact with their environment. In particular, it is argued that only states that are not modified by such an interaction are stable and, therefore, observable. More formally, the preferred basis $\{ |\psi_i\rangle \}$ is supposed to be the one that satisfies
\begin{equation}
|\psi_i\rangle |E_0\rangle \xrightarrow{\text{Schrödinger}} |\psi_i\rangle |E_i\rangle
\end{equation}
with $ \langle E_i | E_j\rangle \approx 0 $, where $|E_0\rangle$ is the initial state of the environment and $|E_k\rangle$ is the state of the environment which results from the fact that the state of the system is $|\psi_k\rangle$. Therefore, if the initial state of the system is a superposition of elements of such a basis,
\begin{equation}
\sum_i c_i  |\psi_i\rangle |E_0\rangle \xrightarrow{\text{Schrödinger}} \sum_i c_i  |\psi_i\rangle |E_i\rangle
\end{equation}
and the system becomes entangled with the environment, supposedly leading to the unobservability of such states. \cite{Oko.Sud:16} describe in detail why all this does not constitute a valid explanation for the fact that we do not seem to observe macroscopic objects on superpositions.\footnote{In a nutshell, the problem is the following. The argument for the suppression of macroscopic interference via decoherence is that, \emph{for all practical purposes}, reduced density matrices of systems in interaction with an environment behave as mixtures. However, those reduced density matrices behave as mixtures \emph{only if} one assumes that, upon measurement, systems collapse \emph{à la} Copenhague. Therefore, in order for the argument to work, one basically needs to assume what one wants to prove (again, see \cite{Oko.Sud:16} for details).}  Here, though, we will employ some elements of the decoherence story, together with an evolutionary perspective, in order to explain why $|S\rangle_O$ never seems to occur.

Suppose that an observer closes her eyes in front of the spin measurement apparatus of section \ref{MP}, while it measures the spin of a spin-up particle. Suppose as well (for now) that everything, always evolves according to the Schrödinger equation. Before the observer opens her eyes, the state of the apparatus and the particle is  $|+\rangle_M|+\rangle_p $. What happens when she opens her eyes? Well, according to what we said in section \ref{MP}, she will end up in the state $|+\rangle_O$ (of course, the analogous thing happens with a spin-down particle).

Now consider the same exercise, but with the apparatus measuring a particle in a superposition of spin-up and spin-down. After the measurement, the state of the apparatus and the particle will be $|S\rangle_{M+p}$, and the important question is what happens when the observer opens her eyes. As we explained in section  \ref{MP}, what we just said about the case where the particle is spin up or spin-down, together with the fact that the Schrödinger equation is linear, completely determine the answer; that is, the fact that
\begin{equation}
\label{hand}
|R\rangle_O |+\rangle_M  | + \rangle_p \xrightarrow{\text{Schrödinger}}  |+\rangle_O |+\rangle_M|+\rangle_p \quad \text{and} \quad |R\rangle_O |-\rangle_M  | - \rangle_p \xrightarrow{\text{Schrödinger}}  |-\rangle_O |-\rangle_M|-\rangle_p
\end{equation}
necessarily implies
\begin{equation}
|R\rangle_O |S\rangle_{M+p}=|R\rangle_O \left\lbrace \alpha |+\rangle_M |+\rangle_p + \beta |-\rangle_M |-\rangle_p \right\rbrace \xrightarrow{\text{Schrödinger}} \alpha |+\rangle_O |+\rangle_M|+\rangle_p + \beta |-\rangle_O |-\rangle_M|-\rangle_p ,
\end{equation}
so the observer will end up in such a superposition, and \emph{not} in the state $|S\rangle_O$. Are we done then? Have we explained why the observer does not end up with a perception of a superposition but a superposition of perceptions (for which we already offered a solution within our model)? Not really. The key point to observe is that the asymmetry between $| \pm \rangle_O$ and $|S\rangle_O$, which is what we are after, has been simply put in by hand by \emph{assuming} that equation (\ref{hand}) is correct. The problem is that, from the fundamental point of view, there is nothing to justify the postulation of such an evolution over the analogue with respect to $|S\rangle_O$, namely
\begin{equation}
\label{handb}
|R\rangle_O \left\lbrace \alpha  |+\rangle_M|+\rangle_p + \beta  |-\rangle_M|-\rangle_p \right\rbrace \xrightarrow{\text{Schrödinger}}  |S\rangle_O \left\lbrace \alpha  |+\rangle_M|+\rangle_p + \beta  |-\rangle_M|-\rangle_p \right\rbrace .
\end{equation}
Of course, we seem to know from \emph{experience} that (\ref{hand}) is reasonable and (\ref{handb}) is not but, as we argued above, a satisfactory solution of the measurement problem requires an explanation of such a fact.

To recap, we made the distinction between a superposition of incompatible perceptions and a perception of a macroscopic superposition. The CSL model we propose takes care of the first scenario by quickly killing such states but it does not help in suppressing the second (note that a position basis CSL model \emph{does} take care of both cases). What we need to do now is to explain why  (\ref{hand}), but not (\ref{handb}), seems to be the case.

We first note that a system that behaves as (\ref{hand}) \emph{can}, at the same time, possess a state $|R^*\rangle_O$ (as opposed to $|R\rangle_O$) that behaves as
\begin{equation}
\label{handbs}
|R^*\rangle_O \left\lbrace \alpha  |+\rangle_M|+\rangle_p + \beta  |-\rangle_M|-\rangle_p \right\rbrace \xrightarrow{\text{Schrödinger}}  |S\rangle_O \left\lbrace \alpha  |+\rangle_M|+\rangle_p + \beta  |-\rangle_M|-\rangle_p \right\rbrace .
\end{equation}
For example, in the same way that a measuring apparatus could, by turning a dial, measure spin along different directions, the brain could have a state $|R\rangle_O$ that behaves as  (\ref{hand}) \emph{and} a different state $|R^*\rangle_O$ that behaves as (\ref{handbs}).

Having noticed that a brain could indeed possess two different ready states that, given the exact same physical situation, would track the world in two different incompatible bases, the key question to ask is if such an architecture would give any edge to a being possessing it. The short answer is that it would not. This is because if those two ready states are indeed available, then a superposition of them also would. But if such a superposition occurs, then the collapse mechanics would \emph{randomly} choose one or the other and only one basis would be tracked, without the conscious being being able to control which. Moreover, it is clear that there is in fact a continuum of bases that could be tracked (not only two, as above), so the randomization problem we mentioned is even worse. So, we can conclude that a brain architecture that tracks systems along a single basis is greatly superior, and hence very likely selected in evolution.

Another way in which a system could, at the same time, behave as (\ref{hand}) and as (\ref{handbs}) is by having within it one module with a state $|R\rangle_O$ that behaves as (\ref{hand}) and another module with a state $|R^*\rangle_O$ that behaves as (\ref{handbs}). That is, one could have a module that ends up either in  $|+\rangle_O$ or  $|-\rangle_O$ as a result of looking at the measuring device \emph{and} a different module that ends up in  $|S\rangle_O$ as a result of the very same interaction. In other words, in principle we could possesses two different modules that track the same property in different bases, such that both inputs contribute to the final experience. The key question, again, is if such an architecture would give any edge to a being possessing it, and the answer, again, is that it would not. That is because, by trying to perceive things with both modules at the same time, no reliable information would be gathered. This is analogous to an apparatus which is supposed to measure at the same time spin along $x$ and $y$ of a particle. Of course, since those two properties do not commute, such an apparatus would end up not giving information with respect to either property.
 
Moreover, this possibility does not even start to make sense in IIT, which we are taking for granted. We should not think of our overall state of consciousness, within the theory, as an aggregation of conscious states. As we have seen, the exclusion postulate states at the system level that, for a system, there is a unique conceptual structure that gives rise to consciousness, and therefore that constellations generated by overlapping elements are excluded: each experience excludes all others. So, we should think of the overall state of consciousness of the observer before looking at the measuring device and such a state can either be $|R^*\rangle_O$ or $|R\rangle_O$ but not both. It is a matter of the internal structure of the cognitive systems whether they are are like one or the other. It turns out that, for the human architecture, it is (\ref{hand}), which describes the interaction. But this is a contingent fact and the theory leaves open the possibility of different cognitive architectures that lead to a conceptual structure that corresponds to initial states like $|R^*\rangle_O$. In this case, their observation of the measuring device would be described by (\ref{handbs}): they would be systems that track superpositions of locations, rather than well-defined positions as we do.

\section{Conclusion}
\label{C}
Many authors since the early days of quantum mechanics have played around with the notion that consciousness holds the key for understanding the collapse of the wave function. However, such an idea seems to suggest that consciousness lies outside of the quantum realm. This fact, together with an absence of a formal model for the alleged interaction between consciousness and the material world, has contributed to the lack of popularity of theories in which ``consciousness causes the collapse of the wave function.''

In this paper we have shown that it is possible to provide such a model. For this purpose, we have turned to IIT, which maintains that consciousness depends upon integration of information of a system and provides a clear way to measure it. However, it is worth stressing once again that our approach does not depend on IIT being the correct theory of consciousness; we have merely employed  it for illustrative purposes. All our model requires is the construction of an operator that measures consciousness. As long as there is a physical property upon which consciousness depends, and which can be measured, then we can use such a property as a collapse operator in order to construct a consciousness-based CSL model.

It is worth noting that the standard choice for a collapse operator, in terms of position, is of course well justified by observations, but lacks an explanation or an independent motivation. In our proposal, in contrast, the fact that we never observe superpositions in the position of macroscopic objects is simply a contingent fact, derived from the cognitive architecture that happens to give rise to consciousness in our case. The collapse term is ``cooked'' to get rid of superpositions of conscious states, but the idea that consciousness ``causes'' the collapse of the wave function is motivated independently.  

Unlike other approaches that have postulated a role for consciousness in the collapse of the wave function, we offer a clear and formal model of such an interaction. We are sure that there are further conceptual questions to be attended, but we hope this model contributes to making these approaches more attractive and to the development of alternative models. One should not forget that it is still an open empirical question whether consciousness is really involved in the collapse of the wave function (see \cite{O+S} for an empirical setup that could answer such a question).

\section*{Acknowledgments}
We are very grateful to David Chalmers, Martin Glazier, Kelvin McQueen and Manolo Martinez for their useful comments. Financial support for this work was provided by DGAPA projects IG100316 and IA400218.
\bibliographystyle{apalike}
\bibliography{biblio,librobib}
\end{document}